\documentclass[aps,12pt,tightenlines,showpacs,showkeys]{revtex4}
\usepackage{graphicx}
\usepackage{dcolumn}
\begin{document}

\title{ The Antiparticles of Neutral Bosons }

\author{W. A. Perkins}

\affiliation {Perkins Advanced Computing Systems,\\ 12303 Hidden Meadows
Circle, Auburn, CA 95603, USA\\E-mail: wperkins@nccn.net} 

\begin{abstract}
With the advent of the ability to create and study antihydrogen, we think it is appropriate to consider the possibility that antiphotons might not be identical to photons. First of all, we will look at the experimental evidence concerning multiple neutral pions and multiple photons. Because of its internal structure, the neutral kaon is not identical to its antiparticle. We will consider internal structures for the neutral pion and photon for which the antiparticle differs from the particle. Interestingly, the antiphoton thus created from neutrinos does not interact with electrons because its neutrinos have the wrong helicity.

\end{abstract}




\maketitle

\section{\label{sec.intro}Introduction}

In the last few years researchers~\cite{andresen1} have succeeded in forming and trapping large numbers of antihydrogen atoms for long periods of time. Scientists~\cite{amole1} in addition have measured resonant quantum transitions of antihydrogen atoms. They have not measured the emission spectrum yet, and the purpose of this paper is to consider the possibility that antiphotons are different than photons.

The ``neutrino theory of light'' provides a natural way for antiphotons to differ from photons. Photons are changed into antiphotons by applying  the charge conjugation operator on the photon's internal neutrino structure. In 1938 Pryce~\cite{pryce1} presented several reasons why de Broglie's idea of a composite photon formed of a neutrino-antineutrino pair, ``neutrino theory of light'', was untenable, and it has had a stifling effect on the theoretical work ever since. However, in 1966, Berezinskii~\cite{berezinskii1} showed that Pryce's only valid argument was that the composite photon could not satisfy Bose commutation relations because of its internal fermion structure. Today, that objection does not seem so important as many composite bosons (such as pions and kaons) cannot satisfy that criterion~\cite{perkins7}.

There has been some continuing work on the neutrino theory of 
light (see~\cite{dvoeglazov1,dvoeglazov2,perkins5}), but it still has other problems: (1) It is unclear how a neutrino-antineutrino pair interacts with an electron with the electromagnetic coupling constant ``$\alpha$'' while a single neutrino interacts with the weak coupling constant ``g'', and how the local interaction between the neutrino and the antineutrino affects this. (2) The theory needs massless 2-component neutrinos, while there are indications~\cite{fukuda} that neutrinos have mass. Nevertheless, in this paper we will consider neutrinos to be massless 2-component spinors~\cite{lee1}.

In the next section we will consider the evidence that multiple neutral bosons exist. The existence of multiple neutral kaons is well established, while the evidence for multiple neutral pions and multiple photons is less well-known. If the reactions $\overline{p}{p} \rightarrow {\pi^0} {\pi^0}$ and $\overline{p}{d} \rightarrow {\pi^0} {\pi^0} n$ occur from S-states as implied by the experimental results, then the two neutral pions are not identical. In addition, there are results supporting the existence of a short-lived $\pi^0$ from antiproton annihilations in emulsions. Indirect evidence of multiple photons comes from the decay $\pi^0 \rightarrow \gamma + \gamma$.

In Section~\ref{sec.theory} we look at internal structures for which the antiparticles are different than the particles. From the ``neutrino theory of light'' the photon is composed of the left-handed electron neutrino and the right-handed electron antineutrino. Under charge conjugation the photon changes into an antiphoton composed of a right-handed electron neutrino and a left-handed electron antineutrino, two particles that have never been observed. Then in Section~\ref{sec.photon_inter} we examine the interaction of composite photons and antiphotons with elections. Because the electron-neutrino interaction is V-A in a matter world, it selects states with negative-helicity particles and positive-helicity antiparticles. An antiphoton cannot interact with electrons because its neutrinos have the wrong helicity. Conversely, a photon cannot interact with positrons in an antimatter world where the interaction is V+A.

\section{\label{sec.evidence} Evidence for Multiple Neutral Bosons}

Neutral fermions, such as neutrinos and the neutron, are not identical with their antiparticles. Although it is expected that neutral bosons will be identical with their antiparticles, the neutral kaon, $K^0$, is known to differ from its antiparticle, $\overline{K}^0$.

\subsection{Multiple Neutral Kaons}

To conserve strangeness in strong reactions, the $K^0$ must have strangeness +1 while the $\overline{K}^0$ must have strangeness -1. One also observes neutral kaons with different lifetimes, ${K^0_S}$ and ${K^0_L}$.

One method to test if two neutral particles are identical is to look for decays which are forbidden for identical particles. If the neutral kaon were identical with its antiparticle, the reaction $\overline{p}{p} \rightarrow {K^0} {K^0}$ could not occur from as S-state of the $\overline{p} {p}$ system. The fact that the reaction occurs readily compared with $\overline{p}{p} \rightarrow {K^+}{K}^-$ and since it has been determined~\cite{bizzarri1} that $\overline{p}{p} \rightarrow {K} {K} < 6\%$ from P states, this is evidence that the neutral kaon differs from its antiparticle.

\subsection{Multiple Neutral Pions}

We will now consider evidence that at least two distinct neutral pions exist. Initially, it was thought that the reaction $\overline{p}{p} \rightarrow {\pi^0} {\pi^0}$ might not be observed since it is forbidden from an S-state of the $\overline{p}{p}$ system if the two $\pi^0$ are identical~\cite{perkins3}. To explain its occurrence for identical $\pi^0$'s requires an anomalously high fraction from P-states as shown in {\bf Table 1}. Theoretically it is expected~\cite{day1,leon1} that absorption will occur predominately from S-states for $\pi^-$ and $K^-$ and for protonium capture~\cite{desai1}. Experimental results for $\pi^- p$~\cite{fields1,bierman1}, 
$K^- p$~\cite{cresti1,knop1} and $\Sigma^- p$~\cite{burnstein1} show predominance of S-state capture.

\begin{table}
\begin{ruledtabular}
\caption{Branching Ratio for $\overline{p} p \rightarrow \pi^0 \pi^0$}
\begin{tabular}{cccc}
Measured value& \% from P states&Year&Reference \\
\hline
$(4.8\pm 1.0)\times 10^{-4}$&$39\%$&$1971$& Devons 
{\it et~al.}~\cite{devons1}\\
$(1.4\pm 0.3)\times 10^{-4}$&$13\%$&$1979$& Bassompierre 
{\it et~al.}~\cite{bassompierre1}\\
$(6\pm 4)\times 10^{-4}$&$47\%$&$1983$& Backenstoss 
{\it et~al.}~\cite{backenstoss1}\\
$(2.06\pm 0.14)\times 10^{-4}$&$18\%$&$1987$& Adiels 
{\it et~al.}~\cite{adiels1}\\
$(2.5\pm 0.3)\times 10^{-4}$&$22\%$&$1988$& Chiba 
{\it et~al.}~\cite{chiba1}\\
$(6.93\pm 0.43)\times 10^{-4}$&$53\%$&$1992$& Crystal Barrel ~\cite{crystal1}\\
$(2.8\pm 0.4)\times 10^{-4}$&$24\%$&$1998$& Obelix~\cite{obelix1}\\
$(6.14\pm 0.40)\times 10^{-4}$&$48\%$&$2001$& Crystal Barrel~\cite{crystal2}\\
\end{tabular}
\end{ruledtabular}
\end{table}

The reaction $\overline{p}{d} \rightarrow {\pi^0} {\pi^0} n$ is also not allowed from S-states if the two $\pi^0$'s are identical. Using charge independence one can calculate the fraction that must occur from the measured branching ratio of the reactions $\overline{p}{d} \rightarrow {\pi^-} {\pi^0} p$ and $\overline{p}{d} \rightarrow {\pi^+} {\pi^-} n$ with the results shown in 
{\bf Table 2} indicating large fractions from P-states.

Efforts~\cite{klempt1,batty1,batty2} to reduce these discrepancies (large fraction captured from P-states when S-state capture should dominate) fall far short~\cite{perkins3}. However, if $\overline{p}{p} \rightarrow {\pi^0} {\pi^0}$ and $\overline{p}{d} \rightarrow {\pi^0} {\pi^0} n$ can occur from S-states the anomaly disappears. Therefore, we conclude that there is strong evidence that multiple $\pi^0$'s exist, and experimental tests to confirm this have been proposed~\cite{perkins3}.

\begin{table}
\begin{ruledtabular}
\caption{Ratio BR($\overline{p} d \rightarrow \pi^- \pi^0 p$)/
BR($\overline{p} d \rightarrow \pi^+ \pi^- n$) }
\begin{tabular}{ccccc}
Measured value& Method& \% from P states&Year&Reference \\
\hline
$(0.68\pm 0.07)$&deut. bub. cham.&$75\%$&$1973$& Gray 
{\it et~al.}~\cite{gray1}\\
$(0.70\pm 0.05)$&mag. spect.&$74\%$&$1986$& Bridges 
{\it et~al.}~\cite{bridges1}\\
$(0.55\pm 0.05)$&mag. spect.&$80\%$&$1986$& Bridges 
{\it et~al.}~\cite{bridges1}\\
$(1.48\pm 0.05)$\footnote{Corrected for pairs beyond 5 degrees, see~\cite{perkins3}}&mag. spect.&$34\%$&$1988$& Angelopoulos 
{\it et~al.}~\cite{angelopoulos1}\\
\end{tabular}
\end{ruledtabular}
\end{table} 

Further grounds for belief in the existence of multiple neutral pions comes the experiments of Tsai-Chu
{\it et~al.}~\cite{tsai1,tsai2} with antiprotons annihilating in emulsions. They reported the properties of a second neutral pion: (1) It has mass of the same order as the usual $\pi^0$, (2) It is emitted with the same energy as that of a charged pion, (3) It decays much more often into electron pairs and into double pairs, (4) The electron pairs from this second neutral pion have larger opening angles than those of Dalitz pairs, and (5) It has a very short lifetime (much shorter than the usual $\pi^0$) because the electrons are emitted directly from the origin of the annihilation stars. One emulsion event~\cite{tsai1}, attributed to an antineutron annihilation, appears to involve an $\eta$ decay into three $\pi^0$'s each of which decay into four electrons.

Since there is no indication of such $\pi^0$ decays from antiproton annihilation in liquid hydrogen, Perkins~\cite{perkins3} concluded that this second neutral pion has a lifetime so short that it occasionally decays before it can leave the annihilation nucleus (e.g., Ag) of the emulsion. The probability for creation of electron pairs by $\pi^0$ photons inside a nucleus is very high, and the opening angles for pairs produced by high-energy photons on nuclei are wider than those for Dalitz pairs. 

It is uncertain whether such a short-lived $\pi^0$ would have been detected in the Primakoff-effect experiments~\cite{bernstein1} because of the high angular resolution required in the forward direction. 

\subsection{Multiple Types of Photons}

We will now consider evidence that at least two distinct photons exist. Again we look at decays that are forbidden if the photons are identical. A vector particle cannot decay into two photons according to the theorem of Landau~\cite{landau} and Yang~\cite{yang}. The two-photon state must be described in terms of three vectors: the relative momentum ${\bf k}$, and the two polarization vectors ${\bf e_1}$ and ${\bf e_2}$. Since the state must be bilinear in the polarization vectors, there are just three possibilities:
\begin{eqnarray}
{\bf e_1} \times {\bf e_2}, \nonumber \\
({\bf e_1} \cdot {\bf e_2}) {\bf k}, \nonumber \\
{\bf k} \times ({\bf e_1} \times {\bf e_2}).
\end{eqnarray}
The last one has zero amplitude, while the first two are antisymmetric under an interchange of the two photons. Identical photons must be symmetric under interchange. This theorem has been used to prove that the neutral pion is not a vector particle, because it decays into two photons. However, if the two photons are not identical, the state of two photons can be antisymmetric under an interchange, and the neutral pion can be a vector particle. Therefore, if it can be shown that the neutral pion is a vector particle, this will prove that the two photons are not identical.

\begin{figure}
\includegraphics[scale=0.4]{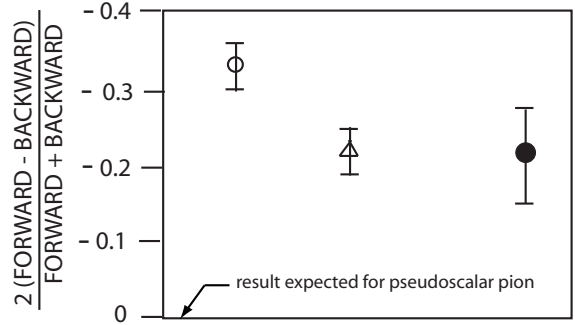}
\caption{Experimental results showing large forward/backward asymmetry in $\pi-\mu$ decay relative to direction of proton beam with more muons emitted in backward direction. $\bigcirc$ Peterson 1950~\cite{peterson}, $\triangle$ Lattes 1957~\cite{lattes}, $\bullet$ Hulubei 1963~\cite{hulubei1}}.
\label{f1}
\end{figure}

If charged pions are vector particles, the neutral pion must also be a vector particle. There is significant evidence from multiple experiments~\cite{perkins4} that the $\pi^+$ carries directional information.
The evidence comes from the asymmetry of muons from $\pi-\mu$ decay at rest. The results of three experiments are shown in {\bf Figure~\ref{f1}}. Notice the large deviation from zero, the expected result for a pseudoscalar pion.
One might expect that this large an asymmetry would be noticeable in many pion experiments. However the muon from $\pi-\mu$ decay at rest only travels about 1 mm in most materials. This problem does not exist for $\pi \rightarrow e + \nu$ decay, but it takes a complex detection apparatus to separate out the background. The muon in $K \rightarrow \mu + \nu$ would have a longer path, but no one has looked for such decay asymmetries.

It has been suggested~\cite{perkins4} that the pion is an helicity-0 vector particle~\cite{perkins4.5}. Such particles are 4-vectors with a net spin of zero. Experiments that can prove that the pion carries directional information have also been proposed~\cite{perkins4}. A positive result from those experiments will not only be evidence that pions are vector particles, but also evidence that multiple photons exist.

\section{\label{sec.theory} Internal Structures}
We will now consider internal structures of neutral bosons in which the antiparticle differs from the particle. Before we can do that, we need to discuss how the different neutrinos will be represented. Here we require that all neutrinos are massless 2-component spinors~\cite{lee1}. We designate the neutrino connected with electrons by $\nu_{e2}$ and its antiparticle by $\overline{\nu}_{e2}$. The subscript ``2'' indicates that the neutrino $\nu_{e2}$ has spin antiparallel to its momentum while $\overline{\nu}_{e2}$ has its spin parallel to its momentum. The annihilation operators for $\nu_{e2}$ and $\overline{\nu}_{e2}$ are $a_2({\bf k})$ and $c_2({\bf k})$ respectively. Similarly, the annihilation operators for $\nu_{e1}$ (neutrino with spin parallel to its momentum) and $\overline{\nu}_{e1}$ are $a_1({\bf k})$ and $c_1({\bf k})$ respectively.
We consider the $\mu^+$ to be the particle and $\mu^-$ to be the antiparticle. (The reason for this choice will become apparent shortly.) Therefore we designate the neutrino connected with muons by $\nu_{1 \mu}$ and its antiparticle by $\overline{\nu}_{1 \mu}$. The subscript ``1'' indicates that the neutrino $\nu_{1 \mu}$ has spin parallel to its momentum while $\overline{\nu}_{1 \mu}$ has its spin antiparallel to its momentum. The annihilation operators for $\nu_{1 \mu}$ and $\overline{\nu}_{1\mu}$ are $b_1({\bf k})$ and $d_1({\bf k})$ respectively. In addition, the annihilation operators for $\nu_{2 \mu}$ and $\overline{\nu}_{2 \mu}$ are $b_2({\bf k})$ and $d_2({\bf k})$ respectively.

Under C (charge conjugation), the neutrino annihilation operator transform as follows:
\begin{eqnarray}
C a_2({\bf k}) = c_1({\bf k}), \nonumber \\
C c_2({\bf k}) = a_1({\bf k}), \nonumber \\
C a_1({\bf k}) = c_2({\bf k}), \nonumber \\
C c_1({\bf k}) = a_2({\bf k}), \nonumber \\
C b_1({\bf k}) = d_2({\bf k}), \nonumber \\
C d_1({\bf k}) = b_2({\bf k}), \nonumber \\
C b_2({\bf k}) = d_1({\bf k}), \nonumber \\
C d_2({\bf k}) = b_1({\bf k}),
\end{eqnarray}
because two-component neutrino fields are not invariant under C.
Under P (parity), the neutrino annihilation operator transform as follows:
\begin{eqnarray}
P a_2({\bf k}) = a_1({\bf -k}), \nonumber \\
P c_2({\bf k}) = c_1({\bf -k}), \nonumber \\
P a_1({\bf k}) = a_2({\bf -k}), \nonumber \\
P c_1({\bf k}) = c_2({\bf -k}), \nonumber \\
P b_1({\bf k}) = b_2({\bf -k}), \nonumber \\
P d_1({\bf k}) = d_2({\bf -k}), \nonumber \\
P b_2({\bf k}) = b_1({\bf -k}), \nonumber \\
P d_2({\bf k}) = d_1({\bf -k}).
\end{eqnarray}

\subsection{Internal Structure of Neutral Kaons}

According to the Standard Model, the neutral kaon consists of a {\it d} quark and an anti-{\it s} quark,
\begin{eqnarray}
K^0 = d \overline{s}, \nonumber \\
\overline{K}^0 = \overline{d}s.
\end{eqnarray}
Note that $\overline{K}^0$ is different than $K^0$ in its internal structure and has strangeness of -1 while $K^0$ has strangeness of +1.
One can have a linear combinations of $K^0$ and $\overline{K}^0$,
\begin{eqnarray}
|K(t)> = \alpha(t)|K^0> + \beta(t)|\overline{K}^0>.
\end{eqnarray}

Forming the usual linear combinations which are eigenstates of CP,
\begin{eqnarray}
|K_S> = {1 \over \sqrt{2}} (|K^0> + |\overline{K}^0>), \nonumber \\
|K_L> = {1 \over \sqrt{2}} (|K^0> - |\overline{K}^0>),
\end{eqnarray}
with $|K_S>$ having CP = +1 and $|K_L>$ having CP = -1. 

\subsection{Internal Structure of Neutral Pions}

According to the Standard Model, the internal structure of the $\pi^0$ consists of {\it u} and {\it d} quarks,
\begin{eqnarray}
\pi^0 = {1 \over{ \sqrt {2}}} (d \overline {d} - u \overline {u}).
\end{eqnarray}
Its antiparticle,
\begin{eqnarray}
\overline{\pi}^0 = {1 \over{ \sqrt {2}}} (\overline{d} d - \overline{u} u),
\end{eqnarray}
is identical to the particle. We are looking for an internal structure for which the antiparticle and particle are different. Since we suspect that the internal structure of the photon involves neutrinos, we will consider neutrinos as an internal structure for pions.

We take the internal structure of the neutral pion to be,
\begin{eqnarray}
\pi^0 = {1 \over {\sqrt{2}} } (\nu_{e2} \overline{\nu}_{\mu 1}
+ \overline \nu_{e2} \nu_{\mu 1})
\label{eqn7ab}
\end{eqnarray}
giving a particle with zero net spin. We chose the $\mu^+$ as the particle to obtain a $\pi^0$ with spin 0 since the internal momentums are in opposite directions. Equation~(\ref{eqn7ab}) does not conserve lepton flavor number, but it does conserves lepton number. The main evidence for conservation of lepton flavor number is the absence of the decay,
\begin{eqnarray}
\mu^+  \rightarrow e^+ + \gamma,
\end{eqnarray}
but this decay is forbidden by lepton number conservation if $\mu^+$ is the particle.

Under the charge conjugation operation,
\begin{eqnarray}
C \nu_{e2} = \overline{\nu}_{e1}, \nonumber \\
C \overline{\nu}_{\mu 1} = \nu_{\mu 2}, \nonumber \\
C \overline{\nu}_{e2} = \nu_{e1}, \nonumber \\
C \nu_{\mu 1} = \overline{\nu}_{\mu 2}.
\end{eqnarray}

Thus the antiparticle of the $\pi^0$ is,
\begin{eqnarray}
\overline{\pi}^0 = {1 \over {\sqrt{2}}} ( \nu_{\mu 2} \overline{\nu}_{e1} + 
\nu_{e1} \overline{\nu}_{\mu 2} )
\end{eqnarray}
Note that not only is the $\overline{\pi}^0$ different than the $\pi^0$, but its neutrinos types are not even considered to exist. (This assumes that the $\mu^+$ is the particle.)

As was done with the kaon, we create linear combinations of the particle and antiparticle,
\begin{eqnarray}
|\pi_S> = {1 \over \sqrt{2}} (|\pi^0> + |\overline{\pi}^0>), \nonumber \\
|\pi_L> = {1 \over \sqrt{2}} (|\pi^0> - |\overline{\pi}^0>),
\end{eqnarray}

Under charge conjugation,
\begin{eqnarray}
C |\pi_S> = |\pi_S>, \nonumber \\
C |\pi_L> = -|\pi_L>, 
\end{eqnarray}
showing that $|\pi_S>$ is an eigenstate of C with value +1,
while $|\pi_L>$ is an eigenstate of C with value -1. Thus, $|\pi_S>$ could be the short-lived neutral pion observed by Tsai-Chu 
{\it et~al.}~\cite{tsai1,tsai2}

Under the combined operation of charge conjugation and parity,
\begin{eqnarray}
CP \pi^0 = {1 \over {\sqrt{2}} } (\overline {\nu}_{e2} \nu_{\mu 1}
+ \nu_{e2} \overline{\nu}_{\mu 1}) = \pi^0, \nonumber \\
CP \overline{\pi}^0 = {1 \over {\sqrt{2}} } (\overline {\nu}_{\mu 2} \nu_{e1}
+ \overline{\nu}_{e1} \nu_{\mu 2}) = \overline{\pi}^0.
\end{eqnarray}
Thus, both $|\pi_S>$ and $|\pi_L>$ are eigenstates of CP with value +1.
The charged pions do not change into each other under charge conjugation.
For example,
\begin{eqnarray}
C (\pi^+ \rightarrow \mu^+ + \overline{\nu}_{\mu 1}) \neq 
(\pi^- \rightarrow \mu^- + \nu_{\mu 2}),
\end{eqnarray}
but
\begin{eqnarray}
CP (\pi^+ \rightarrow \mu^+ + \overline{\nu}_{\mu1}) = 
(\pi^- \rightarrow \mu^- + \nu_{\mu 1}).
\end{eqnarray}
The $\pi^+, \pi^-$ and their antiparticles have internal structures similar to the $\pi^0$ and $\overline{\pi}^0$.

\subsection{Internal Structure of Photons}

Next we take the internal structure of the photon to be,
\begin{eqnarray}
\gamma = \nu_{2e} \overline{\nu}_{2e}
\end{eqnarray}
giving a particle with helicity +1 or -1.
The photon field is~\cite{perkins5},
\begin{eqnarray}
A_\mu(x) =  \sum_{\bf p} {1 \over 2\sqrt{ V \omega_p}}\left\{ 
\left[G_R({\bf p}) {\overline u}^{+1}_{-1}({\bf p}) \gamma_{\mu} 
u^{-1}_{+1}({\bf p})
+ G_L({\bf p}) {\overline u}^{-1}_{+1}({\bf p}) \gamma_{\mu} 
u^{+1}_{-1}({\bf p}) 
\right]e^{i p x} \right. \nonumber \\
\left. + \left[G_R^\dagger({\bf p}) 
{\overline u}^{-1}_{+1}({\bf p}) \gamma_{\mu} u^{+1}_{-1}({\bf p})
+ G_L^\dagger({\bf p}) {\overline u}^{+1}_{-1}({\bf p}) \gamma_{\mu} 
u^{-1}_{+1}({\bf p}) 
\right]e^{-i p x}  \right\},
\label{eqn17a}
\end{eqnarray}
with the annihilation operators for left-circularly and right-circularly polarized photons with momentum ${\bf p}$ given by,
\begin{eqnarray}
G_L( {\bf p}) = {1 \over \sqrt{2}} \sum_{\bf k} F^\dagger( {\bf k}) 
c_2( {\bf -k}) a_2( {\bf p} + {\bf k})  \nonumber \\
G_R( {\bf p}) = {1 \over \sqrt{2}} \sum_{\bf k} F^\dagger( {\bf k}) 
c_2( {\bf p} + {\bf k}) a_2( {\bf -k}),   
\end{eqnarray}
where $F({\bf k})$ is a spectral function. The plane wave neutrino spinors are 
\begin{eqnarray}
u^{+1}_{+1}({\bf p}) = \sqrt{ {E + p_3} \over 2 E} 
\left( \begin{array}{c}
1 \\ {{p_1 + i p_2} \over {E + p_3}} \\
0 \\ 0
\end{array} \right), 
\nonumber \\
u^{-1}_{-1}({\bf p}) = \sqrt{ {E + p_3} \over 2 E} 
\left( \begin{array}{c}
{{-p_1 + i p_2} \over {E + p_3}} \\ 1 \\
0 \\ 0 
\end{array} \right), 
\nonumber \\
u^{-1}_{+1}({\bf p}) = \sqrt{ {E + p_3} \over 2 E} 
\left( \begin{array}{c}
0 \\ 0 \\
1 \\ {{p_1 + i p_2} \over {E + p_3}} \\
\end{array} \right),
\nonumber \\
u^{+1}_{-1}({\bf p}) = \sqrt{ {E + p_3} \over 2 E} 
\left( \begin{array}{c}
0 \\ 0 \\  {{-p_1 + i p_2} \over {E + p_3}} \\ 1
\end{array} \right), 
\label{eqn5ab}
\end{eqnarray}
where $p_\mu = ({\bf p},iE)$, and the superscripts and subscripts on $u$ refer to the energy and helicity states respectively.

Under the charge conjugation operation,
\begin{eqnarray}
C \nu_{2e} = \overline{\nu}_{1e}, \nonumber \\
C \overline{\nu}_{2e} = \nu_{1e}.
\end{eqnarray}

Thus the antiparticle of the photon is, 
\begin{eqnarray}
\overline{\gamma} = \nu_{1e} \overline{\nu}_{1e}.
\end{eqnarray}
Here again, not only is $\overline{\gamma}$ different than $\gamma$, but its neutrinos types are not considered to exist. The antiphoton field is,
\begin{eqnarray}
{\overline A}_\mu(x) =  \sum_{\bf p} {1 \over 2\sqrt{ V \omega_p}}\left\{ 
\left[{\overline G}_R({\bf p}) 
{\overline u}^{-1}_{-1}({\bf p}) \gamma_{\mu} 
u^{+1}_{+1}({\bf p})
+ {\overline G}_L({\bf p}) {\overline u}^{+1}_{+1}({\bf p}) \gamma_{\mu} 
u^{-1}_{-1}({\bf p}) 
\right]e^{i p x} \right. \nonumber \\
\left. + \left[{\overline G}_R^\dagger({\bf p}) 
{\overline u}^{+1}_{+1}({\bf p}) \gamma_{\mu} u^{-1}_{-1}({\bf p})
+ {\overline G}_L^\dagger({\bf p}) {\overline u}^{-1}_{-1}({\bf p}) \gamma_{\mu} 
u^{+1}_{+1}({\bf p}) 
\right]e^{-i p x}  \right\},
\label{eqn17ac}
\end{eqnarray}
with the annihilation operators for left-circularly and right-circularly polarized antiphotons with momentum ${\bf p}$ given by,
\begin{eqnarray}
{\overline G}_L( {\bf p}) = {1 \over \sqrt{2}} \sum_{\bf k} F^\dagger( {\bf k}) 
c_1( {\bf p} + {\bf k}) a_1({\bf -k})  \nonumber \\
{\overline G}_R( {\bf p}) = {1 \over \sqrt{2}} \sum_{\bf k} F^\dagger( {\bf k}) 
c_1({\bf -k}) a_1( {\bf p} + {\bf k}),   
\end{eqnarray}

As was done with the neutral kaon and neutral pion, we create linear combinations of the particle and antiparticle,
\begin{eqnarray}
|\gamma_1> = {1 \over \sqrt{2}} (|\gamma> + |\overline{\gamma}>) \nonumber \\
|\gamma_2> = {1 \over \sqrt{2}} (|\gamma> - |\overline{\gamma}>),
\end{eqnarray}
although the labels ``1'' and ``2'' cannot refer to lifetime since the photon does not decay. Under charge conjugation,
\begin{eqnarray}
C |\gamma_1> = |\gamma_1>, \nonumber \\
C |\gamma_2> = -|\gamma_2>, 
\end{eqnarray}
showing that $|\gamma_1>$ is an eigenstate of C with value +1,
while $|\gamma_2>$ is an eigenstate of C with value -1.
Under the combined operation of charge conjugation and parity,
\begin{eqnarray}
CP \gamma = \overline{\nu}_{2e} \nu_{2e} = \gamma, \nonumber \\
CP \overline{\gamma} = \overline{\nu}_{1e}\nu_{1e} = \overline{\gamma}.
\end{eqnarray}

In previous versions of the ``neutrino theory of light''~\cite{perkins5} the photon was taken to be $\gamma = \nu_{1e} \overline{\nu}_{1e} + \nu_{2e} \overline{\nu}_{2e}$. Eliminating the $\nu_{1e} \overline{\nu}_{1e}$ term is desirable since those neutrinos have not been observed. However, that term was useful in that it allowed the electromagnetic field to transform in the usual way under the operations of parity and charge conjugation~\cite{perkins6}. Now, the electromagnetic field transforms in the usual way only under the combined operation of CP.
 
\section{\label{sec.photon_inter}Interactions of Composite Photons}
The Lagrangian for the photon and electron fields is,
\begin{eqnarray}
{\cal L}= -{1 \over 4} \left( {\partial A_\mu \over \partial x_\nu}
- {\partial A_\nu \over \partial x_\mu} \right)
 \left( {\partial A_\mu \over \partial x_\nu}
- {\partial A_\nu \over \partial x_\mu} \right)
+ i {\overline \Psi_e} \gamma_{\mu} {\partial \Psi_e \over \partial x_\mu}
 - m_e {\overline \Psi_e} \Psi_e 
 - e {\overline \Psi_e} \gamma_{\mu} A_{\mu} \Psi_e. 
 \label{eqn43ab}
\end{eqnarray}

The interaction between a photon and an electron is given by the term,
\begin{eqnarray}
{\cal L}_{int} = -e {{\overline \Psi}_e} \gamma_\mu A_\mu \Psi_e. 
\label{eqn17}
\end{eqnarray}
From Equation~(\ref{eqn17a}) we see that ${\cal L}_{int}$ contains terms involving neutrino spinors, such as
\begin{eqnarray}
 -e {{\overline \Psi}_e} \gamma_\mu  {\overline u}^{+1}_{-1}({\bf p}) \gamma_{\mu} u^{-1}_{+1}({\bf p})\Psi_e, 
\label{eqn40e}
\end{eqnarray}
and 
\begin{eqnarray}
 -e {{\overline \Psi}_e} \gamma_\mu  {\overline u}^{-1}_{+1}({\bf p}) \gamma_{\mu} u^{+1}_{-1}({\bf p})\Psi_e. 
\label{eqn40bx}
\end{eqnarray}

Since the electron-neutrino interaction is V-A, we must insert the projection operator, ${1\over{2}}(1 - \gamma_5)$ to select states with negative-helicity particles and positive-helicity antiparticles. With this insertion the components become,
\begin{eqnarray}
 -e [{{\overline \Psi}_e} \gamma_\mu (1 - \gamma_5) u^{-1}_{+1}({\bf p})] 
[{\overline u}^{+1}_{-1}({\bf p}) \gamma_{\mu} (1 - \gamma_5) \Psi_e], 
\label{eqn40cx}
\end{eqnarray}
and 
\begin{eqnarray}
 -e [{{\overline \Psi}_e} \gamma_\mu (1 - \gamma_5) u^{+1}_{-1}({\bf p})]  
[{\overline u}^{-1}_{+1}({\bf p}) \gamma_{\mu}(1 - \gamma_5) \Psi_e]. 
\label{eqn40d}
\end{eqnarray}
 
Since $ u^{+1}_{-1}({\bf p})$ designates a negative-helicity particle and 
$u^{-1}_{+1}({\bf p})$ a positive-helicity antiparticle the insertion of ${1\over{2}}(1 - \gamma_5)$ does not change the result as, 
\begin{eqnarray}
{1 \over 2}(1 - \gamma_5) u^{-1}_{+1}({\bf p})
= \left( \begin{array}{cccc}
0 & 0 & 0 & 0 \\
0 & 0 & 0 & 0 \\ 
0 & 0 & 1 & 0 \\
0 & 0 & 0 & 1
\end{array} \right)
\sqrt{ {E + p_3} \over 2 E} 
\left( \begin{array}{c}
0 \\ 0 \\
1 \\ {{p_1 + i p_2} \over {E + p_3}} \\
\end{array} \right)
= \sqrt{ {E + p_3} \over 2 E} 
\left( \begin{array}{c}
0 \\ 0 \\
1 \\ {{p_1 + i p_2} \over {E + p_3}} \\
\end{array} \right) 
\end{eqnarray}

However inserting Equation (24), we see that the interaction of an antiphoton with an electron contains components,
\begin{eqnarray}
 -e [{{\overline \Psi}_e} \gamma_\mu (1 - \gamma_5) u^{+1}_{+1}({\bf p})] 
[{\overline u}^{-1}_{-1}({\bf p}) \gamma_{\mu} (1 - \gamma_5) \Psi_e], 
\label{eqn40f}
\end{eqnarray}
and 
\begin{eqnarray}
 -e [{{\overline \Psi}_e} \gamma_\mu (1 - \gamma_5) u^{-1}_{-1}({\bf p})]  
[{\overline u}^{+1}_{+1}({\bf p}) \gamma_{\mu}(1 - \gamma_5) \Psi_e]. 
\label{eqn40b}
\end{eqnarray}
In this case, $(1 - \gamma_5) u^{+1}_{+1}({\bf p}) = 0$, and
$(1 - \gamma_5) u^{-1}_{-1}({\bf p}) = 0$, because $\nu_{1e}$ and ${\overline \nu}_{1e}$ have the wrong helicity, indicating that 
antiphotons do NOT interact with elections in a matter world.
In an antimatter world the interaction of an antiphoton with a positron has terms containing,
\begin{eqnarray}
 -e [{\overline \Psi}_{e^+} \gamma_\mu (1 + \gamma_5) u^{+1}_{+1}({\bf p})] 
[{\overline u}^{-1}_{-1}({\bf p}) \gamma_{\mu} (1 + \gamma_5) \Psi_{e^+}], 
\label{eqn40c}
\end{eqnarray}
and 
\begin{eqnarray}
 -e [{\overline \Psi}_{e^+} \gamma_\mu (1 + \gamma_5) u^{-1}_{-1}({\bf p})]  
[{\overline u}^{+1}_{+1}({\bf p}) \gamma_{\mu}(1 + \gamma_5) \Psi_{e^+}]. 
\label{eqn40bc}
\end{eqnarray} 
Here the positron-neutrino interaction is V+A and ${1\over{2}}(1 + \gamma_5)$ selects states with positive-helicity particles and negative-helicity antiparticles.
If we replace the antiphoton with a photon the components contain,
\begin{eqnarray}
 -e [{\overline \Psi}_{e^+} \gamma_\mu (1 + \gamma_5) 
u^{-1}_{+1}({\bf p})] 
[{\overline u}^{+1}_{-1}({\bf p}) \gamma_{\mu} (1 + \gamma_5) \Psi_{e^+}], 
\label{eqn40ca}
\end{eqnarray}
and 
\begin{eqnarray}
 -e [{\overline \Psi}_{e^+} \gamma_\mu (1 + \gamma_5) 
u^{+1}_{-1}({\bf p})]  
[{\overline u}^{-1}_{+1}({\bf p}) \gamma_{\mu}(1 + \gamma_5) \Psi_{e^+}]. 
\label{eqn40bca}
\end{eqnarray} 
Here $(1 + \gamma_5) u^{-1}_{+1}({\bf p}) = 0$, and
$(1 + \gamma_5) u^{+1}_{-1}({\bf p}) = 0$, because $\nu_{2e}$ and ${\overline \nu}_{2e}$ have the wrong helicity, indicating that photons do not interact with positrons in an antimatter world.

The search for invisible decays of positronium~\cite{badertscher} show that all the photons are visible to a high degree of accuracy. If the antiphoton, $\overline{ \gamma}$, were present it would not be detected. Therefore, those photons must be linear combinations of $\gamma$ and $\overline{\gamma}$, {\it i.~e.}, $|\gamma_1>$ and $|\gamma_2>$.

We know that the interaction of positrons with the electromagnetic field is similar to that of electrons. Even the interaction of antihydrogen with the electromagnetic field is similar to that of hydrogen~\cite{amole1}. This is all right if the effect of virtual photons is the same in matter and antimatter worlds.

\section{\label{sec.conclude}Conclusions}
In order to explain the experimental results indicating multiple neutral pions and indirect evidence of multiple photons, we have had to deviate from the Standard Model with a different internal structure for pions and propose a photon with an internal structure as in 
``the neutrino theory of light''~\cite{perkins5}.

This new theory requires neutrinos with zero mass. Although it has not been determined directly that neutrinos have mass, the interpretation of the SuperKamiokande neutrino experiments~\cite{fukuda} indicates that neutrinos have mass and therefore are not two-component spinors. 

Proposed experiments~\cite{perkins3} to test for a second neutral pion can determine if the suggested deviations from the Standard Model are justified.
Also an experiment to test if charged pions carry directional information has been proposed~\cite{perkins4}.

An important test of these ideas will occur when the photons from antihydrogen are examined. The present theory predicts that the antiphotons from 
antihydrogen will have the wrong helicity for interaction with electrons, and thus the antiphotons will not be detectable. Furthermore, ordinary photons have the wrong helicity for interaction with antihydrogen.
 
\section{\label{sec.acknow} Acknowledgments}

Helpful discussions with Prof.~J.~E.~Kiskis are gratefully acknowledged.


\end{document}